\documentclass[conference]{IEEEtran}
\usepackage{cite}
\usepackage{amsmath,amssymb,amsfonts}
\usepackage{bm}
\usepackage{algorithmic}
\usepackage{graphicx}
\usepackage[caption=false,font=footnotesize]{subfig}
\usepackage{textcomp}
\usepackage{xcolor}
\usepackage{stfloats}
\def\BibTeX{{\rm B\kern-.05em{\sc i\kern-.025em b}\kern-.08em
    T\kern-.1667em\lower.7ex\hbox{E}\kern-.125emX}}
\begin{document}

\title{Energy-efficient End-to-End Communication System with Spiking Neural Networks\\
}

\author{
\IEEEauthorblockN{
Pooja Plavilayil Anil\IEEEauthorrefmark{1},
Te-Chuan Chiu\IEEEauthorrefmark{2},
and Nguyen Van Huynh\IEEEauthorrefmark{1}
}
\IEEEauthorblockA{
\IEEEauthorrefmark{1}School of Computer Science and Informatics,
University of Liverpool, UK\\
\IEEEauthorrefmark{2}Department of Computer Science,
National Tsing Hua University, Taiwan
}
}

\maketitle

\begin{abstract}
2030 is expected to witness the first commercial 6G systems. However, traditional modular communication architectures cannot meet the latency, optimisation, and integration requirements demanded by various sub-millisecond, data-intensive emerging 6G applications. End-to-end (E2E) communication addresses these limitations by replacing traditional processing blocks with neural networks. Nevertheless, conventional artificial neural networks (ANNs) in E2E systems use dense computation. This results in substantial energy consumption and computational resource requirements, which may not be feasible in resource constrained devices. This paper addresses the aforementioned challenge by developing a spiking neural network (SNN) encoder that targets lower encoder inference energy while maintaining communication reliability. In particular, the SNN encoder combines sparse spike activity, sinusoidal temporal modulation, and weighted temporal aggregation to map one-hot encoded messages to distinguishable transmit signals. The model is evaluated across antenna configurations and SNN timesteps to investigate the bit error rate (BER)-energy trade-off. The simulation results show that the SNN encoder achieves comparable BER under suitable timestep selections, while reducing the average estimated encoder energy by 68.93\%, 52.20\%, and 35.69\% for SNN timesteps 2, 3, and 4, respectively, across five antenna configurations. These findings highlight the potential of the proposed SNN encoder for energy-efficient E2E communication systems.
 
\end{abstract}


\section{Introduction}
\begin{figure}[!tb]
    \centering
    \includegraphics[width=\columnwidth,trim=1cm 4.5cm 2cm 1cm,
        clip]{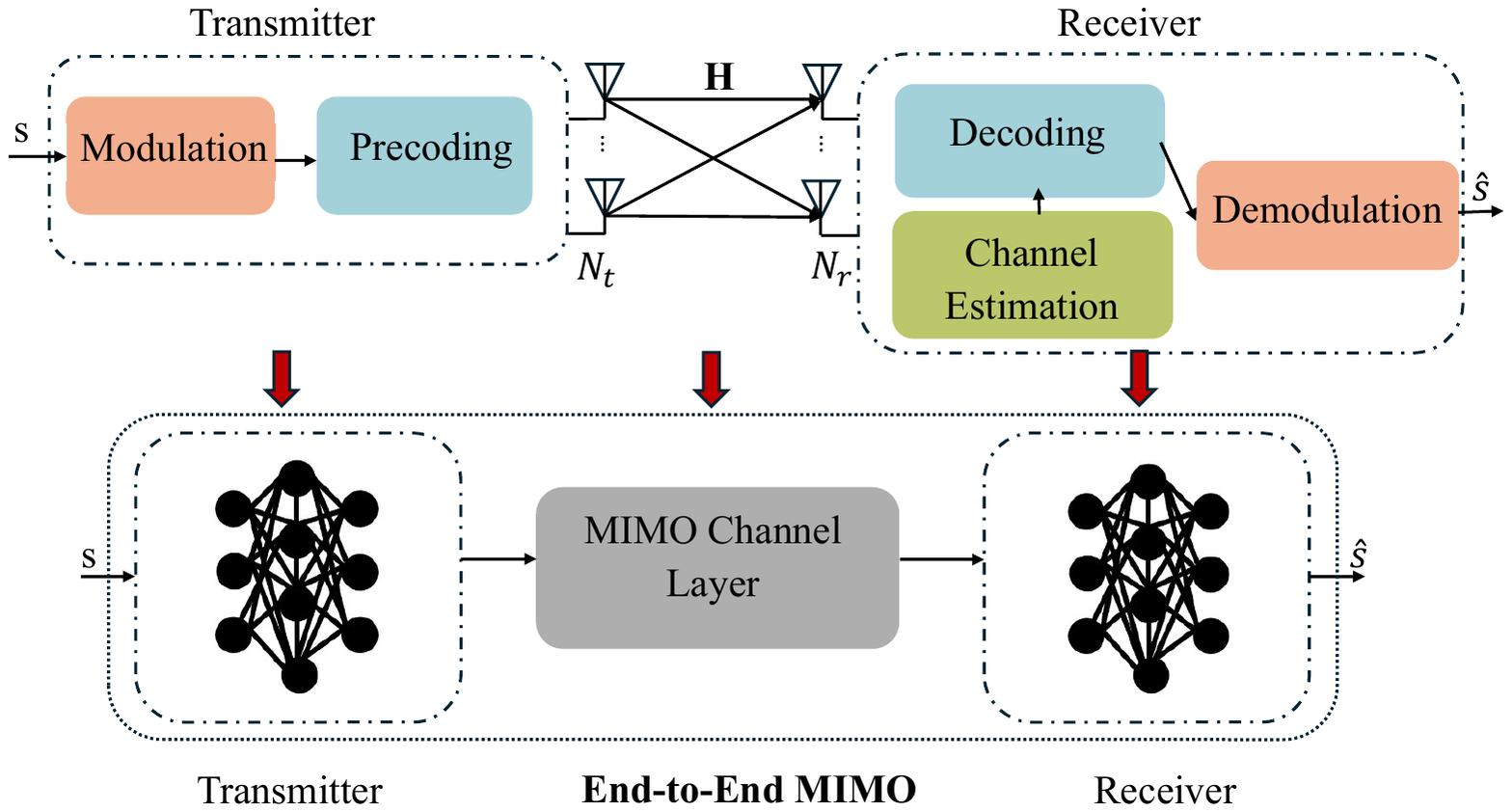}
    \caption{End-to-end MIMO communication. Individual blocks in the traditional communication system are replaced by neural networks at the transmitter and receiver, which are jointly trained.}
    \label{fig:e2e_mimo}
\end{figure}

The evolution towards 6G is expected to enable several emerging data-intensive applications, thus imposing stringent requirements for latency and data rates \cite{Chowdhury20196G}. 6G aims to employ higher-dimensional multiple-input multiple-output (MIMO) systems to meet these performance requirements \cite{huo2023technology}. A conventional MIMO system consists of several signal processing blocks, as illustrated in Fig.~\ref{fig:e2e_mimo}. However, optimising individual blocks separately may limit overall system performance. To address this, end-to-end (E2E) learning \cite{OShea2017AnIT} uses learned transmitter and receiver, as illustrated in Fig.~\ref{fig:e2e_mimo}, enabling joint optimisation under a common communication objective \cite{12Chafi2023}. For example, \cite{DDPG-E2EZhang2025} demonstrates joint training of neural transmitters and receivers using reward feedback without prior channel knowledge, achieving better performance than the evaluated alternatives. The feasibility of practical implementation of real-time artificial neural network (ANN) based E2E receiver processing is also demonstrated in \cite{ DesignWiesmayr2025}, where a standard-compliant 5G New Radio (NR) neural receiver achieves inference latency below 1 ms.

Along with latency and detection performance, the energy per inference is also an important consideration in E2E communication, as larger MIMO configurations can increase neural computation and memory access costs. Specifically, conventional dense ANNs perform multiply-and-accumulate (MAC) operations across their connections, thus increasing energy consumption. This motivates the investigation of an energy-efficient alternative neural network for the E2E communication system. In this context, spiking neural networks (SNNs) represent and process information using discrete events called spikes, where the information is encoded through spike timing and firing rates. As spike activity is typically sparse and spike triggered synaptic computations use accumulate (AC) operations rather than MAC, SNNs can offer potential energy-efficiency advantages \cite{9395703}.

Several studies have investigated SNNs for communication systems, especially at the receiver, demonstrating promising error rate performance and energy savings. The hybrid SNN transformer receiver in \cite{SpikeTimeRxMANS2025} achieves performance close to that of a receiver with perfect channel state information (CSI). The hybrid ANN–SNN receiver in \cite{NeuromorphicRxGupta2026} jointly performs channel estimation, equalisation, and demapping, achieving performance comparable to ANN-based receivers while consuming less energy. Beyond simulation, \cite{ DNN-SNNLiu2024} demonstrates the detection of MIMO-OFDM symbols using a model implemented on Loihi. The results show lower energy per sample while maintaining an accuracy comparable to the evaluated deep neural network (DNN) models. In addition to receiver processing, jointly trained SNN encoders and decoders are developed for remote classification of sensor information and remote inference using semantic communications in \cite{ E2ESkatchkovsky2020} and \cite{ NeuromorphicChen2023}, respectively. 

While the neuromorphic receiver investigates detection and receiver processing, the cited E2E neuromorphic models study task-oriented inference. Their objectives hence differ from transmitter side encoding for reliable message recovery. This motivates investigating whether an SNN encoder can maintain BER comparable to an ANN encoder, while reducing the encoder energy. However, this requires introducing temporal structure into non-temporal one-hot encoded messages. In addition, spike outputs need to be converted into continuous valued transmit signals, while reducing spike activity and retaining distinguishable message representations. To address these critical challenges, the proposed SNN encoder uses sinusoidal temporal modulation, weighted temporal aggregation, and spike penalty. Temporal modulation introduces time-dependent currents, weighted temporal aggregation combines outputs across timesteps to form continuous-valued transmit signals, and the spike penalty encourages sparse spike activity. Experiments across different SNN timesteps and antenna configurations demonstrate the trade-off between BER and the estimated encoder inference energy. The results show that, with appropriate timestep selection for each antenna configuration, the proposed encoder achieves BER comparable to or lower than the conventional ANN encoder while consuming less estimated encoder inference energy.

\section{End-to-End MIMO Communication Model}

Let $\mathcal{M}=\{m_\mathrm{1},m_\mathrm{2},\ldots,m_\mathrm{M}\}$ denote the set of $M$ possible messages, and $\mathcal{S}={\bm{s}_\mathrm{1},\bm{s}_\mathrm{2},\ldots,\bm{s}_\mathrm{M}}$ denote the corresponding set of one-hot encoded vectors. In an E2E MIMO system, the transmitter DNN maps a message to a complex transmit signal matrix according to the mapping, $f_{\bm{\theta}_\mathrm{t}} : \mathcal{S} \rightarrow \mathbb{C}^{N_\mathrm{t} \times \mathrm{T}},$ where $\bm{\theta}_\mathrm{t}$ represents the encoder parameters, $N_\mathrm{t}$ is the number of transmitter antennas and $\mathrm{T}$ denotes the number of channel uses. The encoded transmitter signal, $\bm{X}_\mathrm{k} \in \mathbb{C}^{N_\mathrm{t}\times T}$, can then be expressed as:
\begin{equation}
    \bm{X}_\mathrm{k} = f_{\mathrm{\theta}_\mathrm{t}}(\bm{s}_\mathrm{k}).
    \label{eq:encoder_function_equation}
\end{equation}

At the receiver, the DNN decoder produces message logits as:
\begin{equation}
    \bm{z}_\mathrm{k} = f_{\mathrm{\theta}_\mathrm{r}} (\bm{Y}_\mathrm{k},\hat{\bm{H}}_\mathrm{k}),
    \label{eq:decoder_function_equation}
\end{equation}
where $\bm{z}_\mathrm{k}\in\mathbb{R}^{M}$ contains the logits corresponding to $M$ possible messages, $\bm{Y}_\mathrm{k} \in \mathbb{C}^{N_\mathrm{r} \times T}$ is the received signal matrix corresponding to the transmitted message \(\bm{m}_\mathrm{k}\), and $\hat{\bm{H}}_\mathrm{k} \in \mathbb{C}^{N_\mathrm{r} \times N_\mathrm{t}}$ is the estimated channel matrix available at the receiver. 
The detected message, $\bm{\hat{m}}_\mathrm{k}$, is then obtained as:
\begin{equation}
    \bm{\hat{m}}_\mathrm{k} = \bm{m}_\mathrm{{\hat i_k}},
    \qquad
    \hat i_\mathrm{k}=\underset{i \in \{1,\ldots,M\}}{\arg\max}\; \bm{z}_\mathrm{{k,i}}.
    \label{eq:detected_message_equation}
\end{equation}

\section{Proposed Framework}
\begin{figure}[!tb]
    \centering
    \includegraphics[
        width=\columnwidth,
        keepaspectratio,
        trim=2.2cm 1.2cm 2.5cm 3.2cm,
        clip
    ]{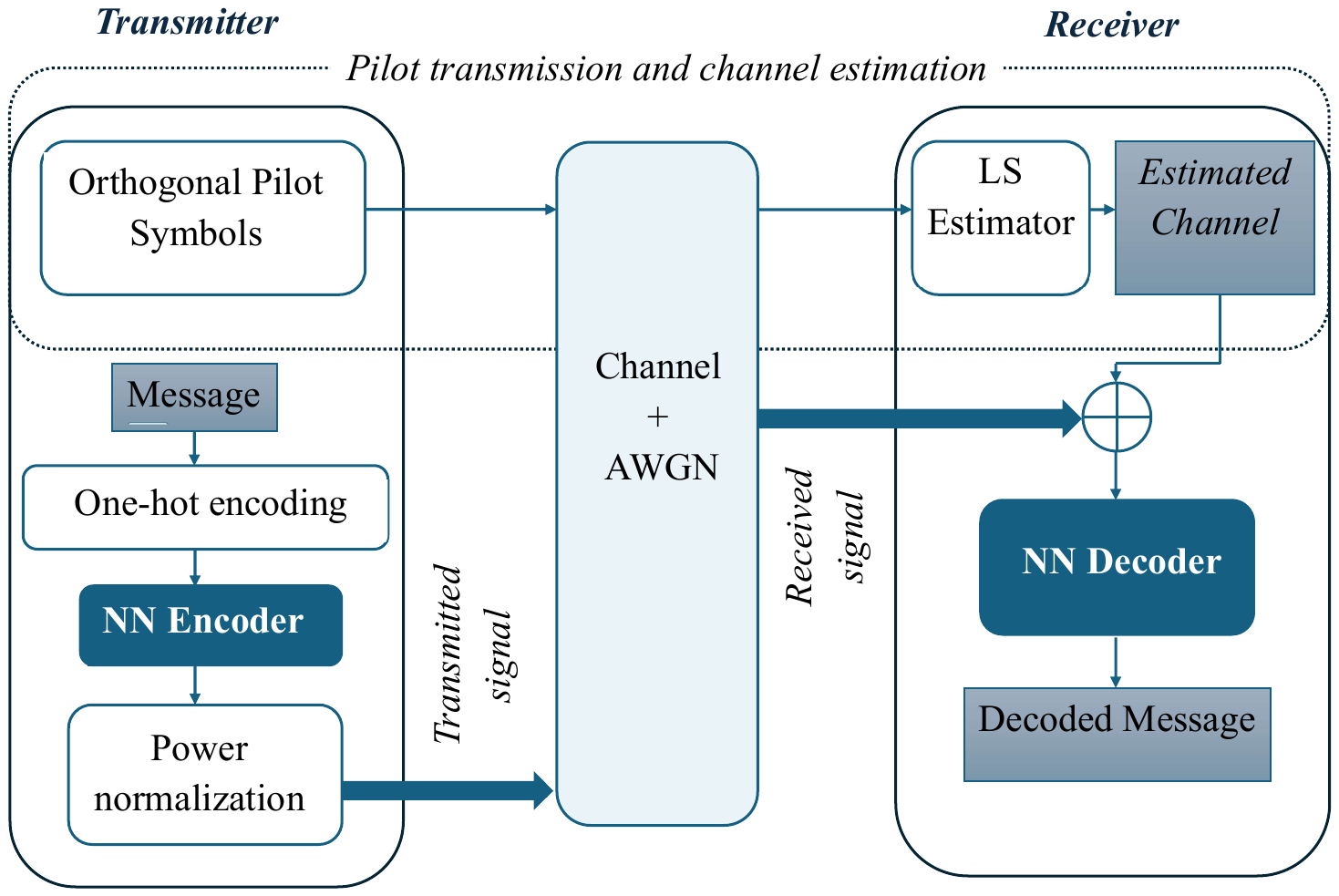}
    \caption{Proposed system model. Orthogonal pilots enable
    LS channel estimation. The encoder maps one-hot messages
    to power-normalised transmit signals, and the decoder
    recovers the messages using the received signals and
    estimated channel.}
    \label{fig:sys_model}
\end{figure}

The framework of the proposed model is illustrated in Fig.~\ref{fig:sys_model}. The channel is modelled as an independent and identically distributed (i.i.d.) Rayleigh fading channel with additive white Gaussian noise (AWGN). As the channel is not known in advance, it is estimated at the receiver using orthogonal pilot signals. If $\mathbf{H}$ is the channel matrix, $\mathbf{X}_\mathrm{p}$ is the transmitted pilot signal, and $\mathbf{N}$ is the AWGN, then the received pilot signal is $\mathbf{Y}_\mathrm{p} = \mathbf{H}\mathbf{X}_\mathrm{p} + \mathbf{N}$. The channel is estimated using the least squares (LS) method \cite{kay1993estimation} and the estimated channel $\hat{\textbf{H}}$ is given by:
\begin{equation}
    \hat{\textbf{H}} = \textbf{Y}_\mathrm{p} \textbf{X}_\mathrm{p}^{H} (\textbf{X}_\mathrm{p} \textbf{X}_\mathrm{p}^{H})^{-1}.
\end{equation}

As shown in Fig.~\ref{fig:sys_model}, the encoder output is normalised to satisfy the transmitter energy constraint as follows:
\begin{equation}
    \bm{x}_\mathrm{k} =
    \frac{\bar{\bm{x}}_\mathrm{k}}
    {\sqrt{
    \frac{1}{N_\mathrm{t}D}
    \left\|\bar{\bm{x}}_\mathrm{k}\right\|_\mathrm{2}^2
    + \epsilon
    }},
    \label{eq:power_normalisation}
\end{equation}
where \(\bar{\bm{x}}_\mathrm{k}\) denotes the unnormalised encoder output, \(D\) is the dimension of the encoder output, and \(\epsilon=10^{-8}\) is a small constant introduced for numerical stability.

For the study, ANN-based and SNN-based encoders are developed. In both cases, decoding is performed using the same ANN decoder architecture. The ANN encoder contains two fully connected layers with 256 and 64 nodes, respectively, each followed by a normalisation layer and ReLU activation function. The output layer has $2N_\mathrm{t}T$ nodes and produces the transmit signal.

The input of the shared ANN decoder consists of flattened received signal $\mathbf{y} \in \mathbb{R}^{2N_\mathrm{r}T}$ concatenated with the estimated CSI, $\hat{\mathbf{H}} \in \mathbb{R}^{2N_\mathrm{r}N_\mathrm{t}}$. The decoder contains three fully connected hidden layers with 128, 64, and 32 nodes, followed by layer normalisation and GeLU activation  function. The output layer has $M$ nodes, which produce logits, $\mathbf{z}\in \mathbb{R}^{B \times M}$, where $B$ is the batch size.

\subsection{Proposed SNN Transmitter}
\begin{figure*}[!t]
    \centering
    \includegraphics[width=\textwidth,
        height=4.2cm,
        keepaspectratio,
        trim=1.5cm 11.2cm 0.7cm 1.2cm,
        clip]{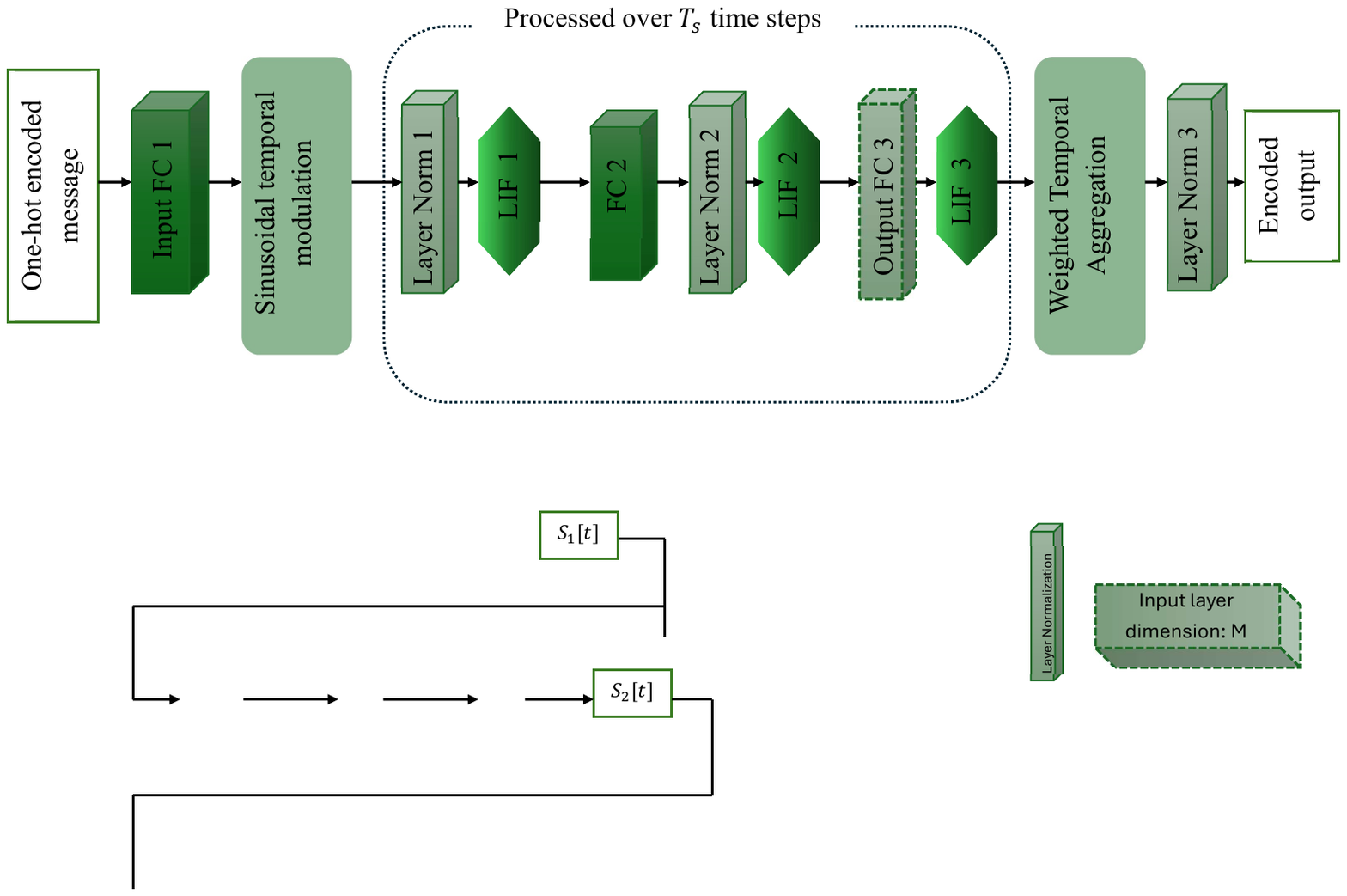}
    \caption{SNN encoder architecture.  A dense input projection and sinusoidal temporal modulation precede fully connected and LIF layers processed over $T_\mathrm{s}$ timesteps, followed by weighted temporal aggregation.}
    \label{fig:snn}
\end{figure*}
For a fair comparison, the SNN and ANN encoders share the same input and network depth. Spiking neurons integrate input currents through membrane potential dynamics \cite{Gerstner_Kistler_Naud_Paninski_2014}, and emit spikes upon reaching a threshold \cite{Yamazaki2022SpikingNN}. As shown in Fig.~\ref{fig:snn}, the one-hot encoded message is passed through the input layer and then through the first hidden layer, which is a fully connected layer (FC1) with 256 neurons. This maps the one-hot input to a static 256 dimensional representation. As this representation is time-invariant, a sinusoidal temporal modulation is applied to introduce temporal variability as in \eqref{eq:Temporal_Modulation}.
\begin{equation}
    \begin{aligned}
        \bm{c}[t]
        &= \left[1 + \mu \sin\left(\frac{2\pi t}{T_\mathrm{s}-1}\right)\right] \\
        &\quad \times \left(\bm{W}_\mathrm{1}\bm{s}_\mathrm{k}+\bm{b}_\mathrm{1}\right), \\
        t &= 0,\ldots,T_\mathrm{s}-1,
    \end{aligned}
    \label{eq:Temporal_Modulation}
\end{equation}
where $\bm{c}[t]$ is the input current at time step $t$, $T_{\mathrm{s}}$ is the number of SNN time steps, and $\mu$, empirically chosen as 0.8, controls the modulation amplitude. $\bm{W}_\mathrm{1}\bm{s}_\mathrm{k}+\bm{b}_\mathrm{1}$ is the static FC1 output.

The resulting time-varying representation passes through normalisation and leaky integrate-and-fire (LIF) neurons \cite{Yang2020Leaky}. This is followed by fully connected layers of 64 and \(2N_\mathrm{t}T\) neurons, with normalisation and LIF processing. All these layers are processed over $T_\mathrm{s}$ timesteps. The output layer membrane potentials, $v[t]$, are aggregated over $T_\mathrm{s}$ so that the later timesteps receive larger weights than the earlier ones, as below:
\begin{equation}
    \bm{x}
    =
    \sum_{t=1}^{T_\mathrm{s}}
    w_\mathrm{t} \bm{v}[t],
    \qquad
    \text{where }
    w_\mathrm{t}
    =
    \frac{t}
    {\sum_{\tau=0}^{T_\mathrm{s}-1} \tau}.
\end{equation}
The aggregated output is then passed through the final normalisation layer. During backpropagation, the SuperSpike surrogate gradient \cite{Zenke2017SuperSpike} approximates the derivative of the spike generation function, with $\alpha=100$ controlling its sharpness around the firing threshold.

\subsection{Problem Formulation}

The transmitter and receiver are jointly trained with the objective of minimising the expected cross-entropy loss, expressed as:
\begin{equation}
\begin{aligned}
\mathcal{L}_\mathrm{{CE}}(\bm{\theta}_\mathrm{t},\bm{\theta}_\mathrm{r})
&=
\mathbb{E}_\mathrm{{m_k,\bm{H}_k}}
\Bigg[
\int
\ell\Big(
f_{\mathrm{\bm{\theta}}_\mathrm{r}}
(\bm{Y}_\mathrm{k},\hat{\bm{H}}_\mathrm{k}),
\bm{m}_\mathrm{k}
\Big) \\
&\quad \times
p\Big(
\bm{Y}_\mathrm{k}
\mid
f_{\bm{\mathrm{\theta}}_\mathrm{t}}(\bm{s}_\mathrm{k}),
\bm{H}_\mathrm{k}
\Big)
\,d\bm{Y}_\mathrm{k}
\Bigg],
\end{aligned}
\end{equation}
where $\ell(\cdot)$ denotes the cross-entropy loss for an individual prediction, $\bm{H}_\mathrm{k}$ is the actual MIMO channel matrix, and $p(\bm{Y}_\mathrm{k} \mid f_{\bm{\mathrm{\theta}}_{\mathrm{t}}}(\bm{s}_\mathrm{k}), \bm{H}_\mathrm{k})$ is the conditional received signal distribution. The integral averages over received signal randomness, while the outer expectation averages over messages and channel realisations. The training objective is then to find the transmitter and receiver parameters that minimise this expected loss.

For the ANN-based encoder, the total loss is given by $\mathcal{L}_\mathrm{{ANN}} = \mathcal{L}_{\mathrm{CE}}$. For the SNN-based encoder, two additional regularisation terms are introduced to encourage sparse spikes and discourage a non-zero mean in the power normalised transmitter output. The total SNN objective becomes:
\begin{equation}
\mathcal{L}_{\mathrm{SNN}}
=
\mathcal{L}_{\mathrm{CE}}
+
\lambda_{\mathrm{sp}}
\mathcal{L}_{\mathrm{sp}}
+
\lambda_{\mathrm{mean}}
\mathcal{L}_{\mathrm{mean}},
\end{equation}
where $ \lambda_{\mathrm{sp}} $ is the spike penalty coefficient and $\lambda_{\mathrm{mean}}$ is the mean output penalty coefficient. Through trial and error, $\lambda_{\mathrm{sp}}$ is empirically chosen as $10^{-4}$ and $\lambda_{\mathrm{mean}}$ as $5 \times 10^{-4}$. $\mathcal{L}_{\mathrm{sp}}$ is the spike penalty, calculated as the mean spike activity over the SNN timesteps and LIF layers, given as:
\begin{equation}
\mathcal{L}_{\mathrm{sp}}
=
\frac{1}{T_\mathrm{s}}
\sum_\mathrm{{t=0}}^{T_\mathrm{s}-1}
\sum_\mathrm{{\ell=1}}^{3}
\bar{a}_\mathrm{{\ell}}[t],
\end{equation}
where \(\bar{a}_\mathrm{{\ell}}[t]\) denotes the mean spike activity of the LIF layer \(\ell\) at the SNN timestep \(t\). $\mathcal{L}_{\mathrm{mean}}$ penalises the batch averaged squared mean of each sample’s transmitter output, encouraging zero-mean transmit signals.
\begin{equation}
\mathcal{L}_{\mathrm{mean}}
=
\frac{1}{B}
\sum_\mathrm{{b=1}}^{B}
\left(
\frac{1}{D}
\sum_{d=1}^{D}
x_{b,d}
\right)^2,
\end{equation}
where $B$ is the batch size, $x_\mathrm{{b,d}}$ is the power normalised transmitter output and $D=2N_\mathrm{t}T$.

\section{Encoder Energy Consumption Model}
For the study, energy per inference is calculated, which is the estimated energy needed to encode one message into a transmit signal. It consists of the computational energy ($E_\mathrm{{comp}}$) and the energy for memory access ($E_\mathrm{{memory}}$) \cite{Lunghi_2025}. The energy model uses the estimates reported in \cite{Davidson2021}, derived from post layout analysis for SpiNNaker2 development in 22FDX technology. The paper reports the energy for one MAC operation ($E_{\mathrm{MAC}}$) as $293\,\mathrm{fJ}$ and the energy per read access from a 2K-word static random-access memory (SRAM) ($E_{\mathrm{SRAM, read}}$) as  $296\, \mathrm{fJ}$. The authors assume that the SRAM write back energy ($E_{\mathrm{SRAM, write}}$) is approximately the same as the read cost and the energy per accumulation operation ($E_{\mathrm{AC}}$) is one-fifth of the MAC energy cost.

\subsection{ANN Encoder Energy}
The ANN encoder energy is estimated from its MAC operations and associated memory accesses. Let the number of MAC operations be denoted by $N_\mathrm{{MAC}}$, input dimension be $M$, hidden-layer sizes be \(N_\mathrm{1},N_\mathrm{2}\), and output dimension be \(N_{\mathrm{out}}\), then the $N_\mathrm{{MAC}}$ is calculated as:
\begin{equation}
    N_{\mathrm{MAC}} = MN_\mathrm{1}+N_\mathrm{1}N_\mathrm{2}+N_\mathrm{2}N_{\mathrm{out}},
\end{equation}
where $N_{\mathrm{out}}=2N_\mathrm{t}T$. Based on \cite{Davidson2021}, each dense connection update energy is modelled as $E_\mathrm{{ANN, op}} = E_\mathrm{{read}} + E_\mathrm{{read}} + E_\mathrm{{MAC}} + E_\mathrm{{write}}$. Hence the ANN energy per inference $E_\mathrm{{ANN}}$ becomes:
\begin{equation}
    E_\mathrm{{ANN}} = N_{\mathrm{MAC}} \times E_\mathrm{{ANN, op}}.
\end{equation}

\subsection{SNN Encoder Energy}
The SNN encoder contains a dense input projection layer followed by spike driven synaptic computation and time-dependent LIF neuron updates. The estimated energy comprises of these three energy contributions. The first SNN layer has a fully connected dense layer before temporal modulation, giving the number of MAC operations in the layer as $N_{\mathrm{MAC,in}}=MN_\mathrm{1}$. The energy for the layer is:
\begin{equation}
    E_{\mathrm{in}} = N_{\mathrm{MAC,in}} E_{\mathrm{ANN,op}}.
\end{equation}

\begin{table}[t]
\centering
\caption{Simulation and Model Parameters}
\label{tab:simulation_parameters}
\footnotesize
\setlength{\tabcolsep}{4pt}
\begin{tabular}{ll}
\hline
\textbf{Parameter} & \textbf{Value} \\
\hline
Message classes, $M$ & 16 \\
Antenna configurations, $N_\mathrm{t} \times N_\mathrm{r}$
    & $1\times1$, $2\times2$, $4\times4$, $8\times8$, $16\times16$ \\
Channel uses, $T$ & 2 \\
Channel model & i.i.d. Rayleigh fading + AWGN \\
Channel estimation & LS \\
Training SNR & $6+\Delta$ dB, $\Delta\sim\mathcal{U}(-1,1)$ \\
Evaluation SNR & $0$--$12$ dB \\
Training epochs & 4000 \\
Batch size & 512 \\
Training samples & $2.048\times10^{6}$ \\
Validation samples & $1.024\times10^{6}$ \\
Test samples/SNR & $1.024\times10^{8}$ \\
ANN encoder & 256--64 \\
SNN encoder & 256--64 \\ 
SNN Timesteps & $T_\mathrm{s} = 2, 3, 4$ \\
SNN $\tau_{\mathrm{mem}}^{-1},\tau_{\mathrm{syn}}^{-1}$
    & 100, 200 \\
SNN threshold, $V_{\mathrm{th}}$ & 0.3 \\
Decoder & 128--64--32 \\
Encoder learning rate & $5\times10^{-4}$ \\
Decoder learning rate & $7\times10^{-4}$ \\
\hline
\end{tabular}
\end{table}
In the subsequent layers, the sparse spikes generated by the LIF neurons drive the synaptic operations and associated memory accesses, enabling AC operations. The energy needed for each synaptic event $(E_\mathrm{{syn,event}})$ can be written as $E_\mathrm{{syn,event}} = 2E_\mathrm{{read}} + E_\mathrm{{AC}} + E_\mathrm{{write}}$ \cite{Davidson2021}. Let $N_\mathrm{{syn}}$ be the average number of synaptic events per inference, $S_\mathrm{1}$ and $S_\mathrm{2}$ be the mean spike counts per inference in LIF layer 1 and LIF layer 2, respectively, across all \(T_\mathrm{s}\) time steps. Then, $N_\mathrm{{syn}} = S_\mathrm{1}N_\mathrm{2} + S_\mathrm{2}N_\mathrm{{out}}$. Hence, the total synaptic energy per inference, $(E_\mathrm{{syn}})$ is:
\begin{equation}
    E_\mathrm{{syn}} = N_\mathrm{{syn}} E_\mathrm{{syn,event}}.
\end{equation}

The author derives the LIF update energy in \cite{Lunghi_2025} as $E_\mathrm{{lif,upd}} = 2E_\mathrm{{AC}} + 2E_\mathrm{{MAC}}$. Let $N_\mathrm{{LIF}}$ be the total number of LIF neurons in all layers of the SNN encoder, then $N_{\mathrm{LIF}}  = N_\mathrm{1}+N_\mathrm{2}+N_{\mathrm{out}}$. Hence, the energy used for LIF update $E_\mathrm{{LIF}}$ is:
\begin{equation}
    E_\mathrm{{LIF}} = N_\mathrm{{LIF}}T_\mathrm{s}E_\mathrm{{lif,upd}}.
\end{equation}

Therefore, the total SNN encoder energy per inference $E_\mathrm{{SNN}}$ is:
\begin{equation}
    E_\mathrm{{SNN}} = E_{\mathrm{in}} + E_\mathrm{{syn}} + E_\mathrm{{LIF}}.
\end{equation}

The energy evaluation therefore represents analytical energy estimates under adopted hardware assumptions rather than direct measurements on neuromorphic hardware. For the estimate, additional SRAM accesses for LIF time-step state updates, bias additions, normalisation, temporal modulation and aggregation, and spike generation costs are omitted. These exclusions motivate hardware measurements to assess actual energy consumption.

\section{Simulation Results}

In this section, we compare the BER and the estimated encoder energy per inference of the proposed SNN-based model against an ANN reference baseline proposed in \cite{OShea2017AnIT}. Both systems have the same depth and employ the same decoder and are evaluated under the same channel conditions. The simulations use Rayleigh fading with AWGN, LS channel estimation and no CSI available at the transmitter. The simulation and training settings are summarised in Table~\ref{tab:simulation_parameters}. The experiments are conducted across $1\times1$, $2\times2$, $4\times4$, $8\times8$, and $16\times16$ antenna configurations and $T_\mathrm{s} = 2, 3, 4$ to examine the BER-energy trade-off in different antenna regimes.

\begin{figure*}[!t]
    \centering
    \begin{minipage}{0.892\textwidth}
        \centering

        \subfloat[$1\times1$\label{fig:ber_1x1}]{%
            \includegraphics[width=0.49\linewidth]{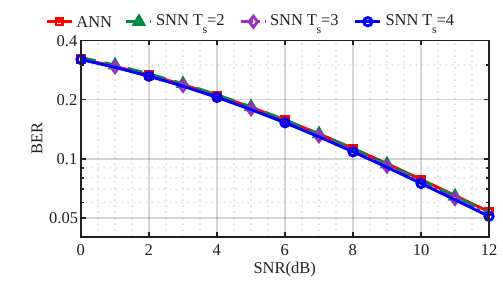}}
        \hfill
        \subfloat[$2\times2$\label{fig:ber_2x2}]{%
            \includegraphics[width=0.49\linewidth]{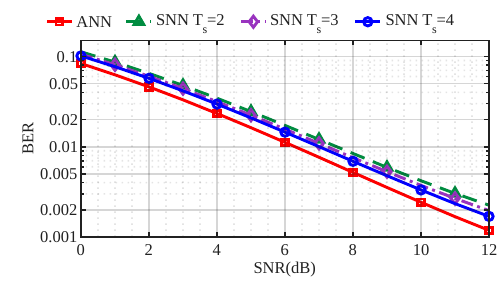}}

        \par\vspace{-2mm}

        \subfloat[$4\times4$\label{fig:ber_4x4}]{%
            \includegraphics[width=0.49\linewidth]{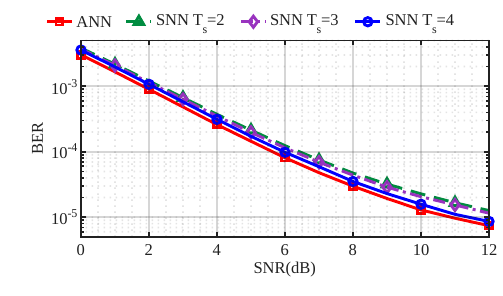}}
        \hfill
        \subfloat[$8\times8$\label{fig:ber_8x8}]{%
            \includegraphics[width=0.49\linewidth]{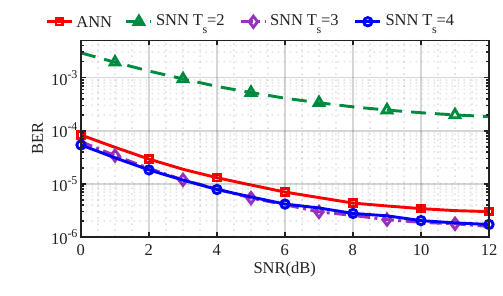}}

        \par\vspace{-2mm}

        \subfloat[$16\times16$\label{fig:ber_16x16}]{%
            \includegraphics[width=0.49\linewidth]{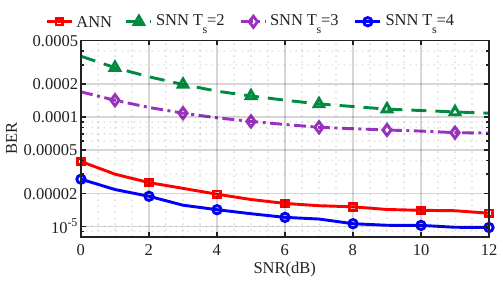}}
    \end{minipage}

    \caption{BER comparison of ANN and SNN encoders
    across SNN timesteps and MIMO antenna configurations.}
    \label{fig:ber_all}
\end{figure*}

\subsection{BER Performance}

The BER performance of the proposed SNN model and its ANN counterpart across MIMO configurations and timesteps is shown in Fig.~\ref{fig:ber_all}. Across all evaluated configurations, the BER decreases consistently with increasing SNR. In \(1\times1\), the SNN achieves a BER close to that of the ANN in all evaluated timesteps. The nearly overlapping curves for $T_\mathrm{s} = 2, 3, 4$ suggest that the limited improvement in BER with additional timesteps may be attributed to the lower dimensionality of the transmitted signal. In \(2\times2\) and \(4\times4\), the ANN has a slightly lower BER than the SNN, across all evaluated SNR range, with only minor improvements with increase in timesteps. Spike activity regularisation in the SNN may constrain the transmit signal representation, whereas the ANN is not subject to this constraint. For \(8\times8\), $T_\mathrm{s} = 3, 4$ achieve similar BER while outperforming the ANN, whereas $T_\mathrm{s} = 2$ causes considerable degradation. The increased spatial dimensionality in the larger antenna configuration and additional temporal integration in higher timesteps may enable the SNN to construct a more favourable arrangement of message representations. In \(16\times16\), although $T_\mathrm{s} = 2, 3$ increases the BER considerably, the SNN achieves lower BER for $T_\mathrm{s} = 4$. This suggests that the additional temporal processing may help the SNN to exploit larger transmit signal dimensions effectively. Overall, these results strongly indicate that the timesteps required to maintain comparable BER depends on the antenna configuration.

\subsection{Encoder Energy Comparison}

\begin{figure*}[!t]
    \centering
    \includegraphics[width=\textwidth]{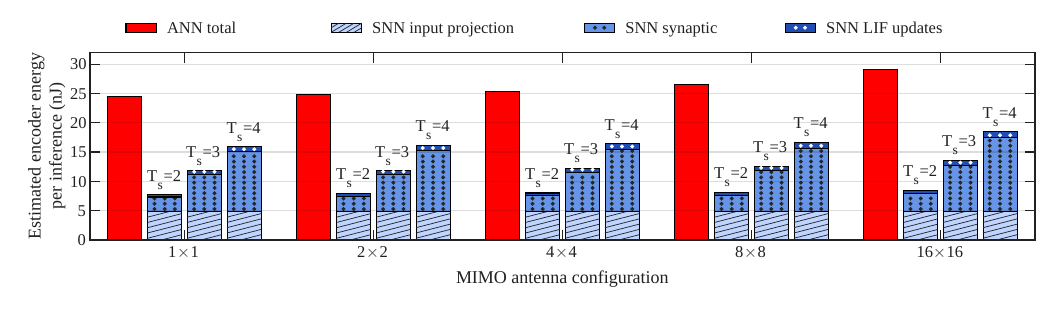}
    \caption{Analytical encoder energy per inference for ANN and SNN across five MIMO antenna configurations, with SNN energy decomposed into dense input projection (FC1), spike-driven synaptic operations, and LIF neuron updates.}
    \label{fig:energy_comparison}
\end{figure*}

The estimated encoder energy per inference of the ANN and SNN transmitters across antenna dimensions and timesteps is compared in Fig.~\ref{fig:energy_comparison}. Across all five antenna configurations, the SNN encoder consumes less energy than its ANN counterpart, with average reductions of 68.93\%, 52.20\%, and 35.69\% for $T_\mathrm{s}=2, 3, 4$, respectively. Also, the SNN energy increases with $T_\mathrm{s}$, as additional timesteps involve further synaptic and LIF operations. In \(1\times1\), $T_\mathrm{s} = 2$ provides the lowest energy with minor BER difference compared to the higher timesteps. In \(8\times8\), $T_\mathrm{s} = 3$ provides a favourable trade-off between energy and BER, as it achieves BER comparable to \(T_\mathrm{s}=4\) with lower estimated energy. For \(16\times16\), reducing \(T_\mathrm{s}\) saves energy but considerably worsens BER, with \(T_\mathrm{s}=4\) retaining both lower BER and lower estimated encoder energy than the ANN. Consequently, the timestep must be selected according to the antenna configuration to preserve BER performance while reducing energy consumption.
It is worth noting that energy accounting only covers encoder inference, excluding training, decoder processing, and channel estimation. The findings are limited to the evaluated SNR range, CSI assumptions, antenna configurations, and training methods.

\section*{Conclusion}
The study investigated the ability of SNN encoders to reduce encoder processing energy while maintaining reliable message recovery. The model uses sinusoidal temporal modulation to represent static messages through time-dependent currents, as well as weighted temporal aggregation to produce continuous valued transmit signals. Across the evaluated antenna configurations, the SNN achieves BER comparable to or lower than its ANN counterpart, while consuming less estimated encoder inference energy. Fewer timesteps lead to greater energy savings, however, the results point to timestep selection as a significant model choice for balancing communication reliability and encoder energy consumption in SNN-based transmitters.

Future work will investigate receiver loss feedback for non-differentiable channels, learned pilots, CSI compression and its availability at the transmitter. Further directions include developing spiking decoders and CSI compressors, investigating realistic channel models, neuromorphic hardware energy validation, and extending energy accounting to the decoder, CSI compressor, and channel estimator.

\bibliographystyle{IEEEtran}
\bibliography{refs}

\end{document}